# Field-like and antidamping spin-orbit torques in as-grown and annealed Ta/CoFeB/MgO layers

Can Onur Avci[1,2], Kevin Garello[1,2], Corneliu Nistor[1,2], Sylvie Godey[2], Belén Ballesteros[2], Aitor Mugarza[2], Alessandro Barla[3], Manuel Valvidares[4], Eric Pellegrin[4], Abhijit Ghosh[1], Ioan Mihai Miron[5], Olivier Boulle[5], Stephane Auffret[5], Gilles Gaudin[5], and Pietro Gambardella[1,2]

[1] *Department of Materials, ETH Zurich, Hönggerbergring 64, CH-8093 Zürich, Switzerland*

[2] *Catalan Institute of Nanoscience and Nanotechnology (ICN2), UAB Campus, E-08193, Barcelona, Spain*

[3] *Istituto di Struttura della Materia (ISM), Consiglio Nazionale delle Ricerche (CNR), S.S. 14 Km 163.5, I-34149 Trieste, Italy*

[4] *ALBA Synchrotron Light Source, E-08290 Cerdanyola del Vallès, Barcelona, Spain*

[5] *SPINTEC, UMR-8191, CEA/CNRS/UJF/GINP, INAC, F-38054 Grenoble, France*

**ABSTRACT**

We present a comprehensive study of the current-induced spin-orbit torques in perpendicularly magnetized Ta/CoFeB/MgO layers. The samples were annealed in steps up to 300ºC and characterized using x-ray absorption spectroscopy, transmission electron microscopy, resistivity, and Hall effect measurements. By performing adiabatic harmonic Hall voltage measurements, we show that the transverse (field-like) and longitudinal (antidamping-like) spin-orbit torques are composed of constant and magnetization-dependent contributions, both of which vary strongly with annealing. Such variations correlate with changes of the saturation magnetization and magnetic anisotropy and are assigned to chemical and structural modifications of the layers. The relative variation of the constant and anisotropic torque terms as a function of annealing temperature is opposite for the field-like and antidamping torques. Measurements of the switching probability using sub-µs current pulses show that the critical current increases with the magnetic anisotropy of the layers, whereas the switching efficiency, measured as the ratio of magnetic anisotropy energy and pulse energy, decreases. The optimal annealing temperature to achieve maximum magnetic anisotropy, saturation magnetization, and switching efficiency is determined to be between 240º and 270 ºC.

I. INTRODUCTION

It has been recently demonstrated that magnetic devices composed of a single ferromagnet (FM) and a heavy metal (HM) layer can be switched by an electric current injected parallel to the plane of the FM.[1] This discovery has attracted considerable interest for applications in spintronics, such as magnetic random access memories.[2-4] Magnetization switching induced by an in-plane current has been investigated mainly in HM/FM/Ox heterostructures, such as Pt/Co/AlOx,[1, 5, 6] Pt/Co/MgO,[7] Ta/CoFeB/MgO,[8-11] W/CoFeB/MgO,[12] and has been shown to work also in all-metal structures such as Pt/Co/Ni,[5] Ta/Ni80Fe20,[13] asymmetric Pt/Co/Pt[14] trilayers and Co/Pd[15] multilayers.

Early work on Pt/Co/AlOx established that a current applied parallel to the interface plane in trilayers with structural inversion asymmetry generates two qualitatively different types of spin torques, namely a transverse (field-like) torque $\mathbf{T}^\perp \sim \mathbf{m} \times \mathbf{y}$[16, 17] and a longitudinal (antidamping-like) torque $\mathbf{T}^\parallel \sim \mathbf{m} \times (\mathbf{y} \times \mathbf{m})$,[1, 5] where $\mathbf{y}$ denotes the in-plane axis perpendicular to the current direction $\mathbf{x}$, and $\mathbf{m}$ is the magnetization unit vector. A recent study[6] further showed that $\mathbf{T}^\perp$ and $\mathbf{T}^\parallel$ include significant additional anisotropic contributions that depend on the magnetization direction relative to the current and symmetry axes, giving the following expressions for the torques:

$$\mathbf{T}^\perp = (\mathbf{y} \times \mathbf{m}) \left[ T_0^\perp + T_2^\perp (\mathbf{z} \times \mathbf{m})^2 + T_4^\perp (\mathbf{z} \times \mathbf{m})^4 \right] +$$

$$\mathbf{m} \times (\mathbf{z} \times \mathbf{m})(\mathbf{m} \cdot \mathbf{x}) \, [T_2^\perp + T_4^\perp (\mathbf{z} \times \mathbf{m})^2] \tag{1}$$

and

$$\mathbf{T}^\parallel = \mathbf{m} \times (\mathbf{y} \times \mathbf{m}) \, T_0^\parallel + (\mathbf{z} \times \mathbf{m})(\mathbf{m} \cdot \mathbf{x}) \, [T_2^\parallel + T_4^\parallel (\mathbf{z} \times \mathbf{m})^2], \tag{2}$$

where the coefficients $T_n^\perp, T_n^\parallel$ represent current-dependent torque amplitudes. The effect of these torques is equivalent to that of two effective fields $\mathbf{B}^\perp = \mathbf{m} \times \mathbf{T}^\perp$ and $\mathbf{B}^\parallel = \mathbf{m} \times \mathbf{T}^\parallel$ perpendicular to the instantaneous magnetization direction, as shown in Fig. 1, which are given by

$$\mathbf{B}^\perp = (\mathbf{y} \times \mathbf{m}) \times \mathbf{m} \, [T_0^\perp + T_2^\perp (\mathbf{z} \times \mathbf{m})^2 + T_4^\perp (\mathbf{z} \times \mathbf{m})^4] +$$
$$(\mathbf{z} \times \mathbf{m})(\mathbf{m} \cdot \mathbf{x}) \, [T_2^\perp + T_4^\perp (\mathbf{z} \times \mathbf{m})^2] \tag{3}$$

and

$$\mathbf{B}^\parallel = (\mathbf{y} \times \mathbf{m}) \, T_0^\parallel + (\mathbf{z} \times \mathbf{m}) \times \mathbf{m}(\mathbf{m} \cdot \mathbf{x}) \, [T_2^\parallel + T_4^\parallel (\mathbf{z} \times \mathbf{m})^2]. \tag{4}$$

These so-called *spin-orbit fields* (or *torques*) originate from the exchange of angular momentum between the crystal lattice and the magnetization, which makes it possible to control the magnetization of a single ferromagnetic layer without transferring spin momentum from a second ferromagnet. The detailed mechanisms giving rise to $\mathbf{T}^\perp$ and $\mathbf{T}^\parallel$ are still debated. Following theoretical work,[18-21] $\mathbf{T}^\perp$ has been attributed to a Rashba-like magnetic field parallel to $\mathbf{y}$[16, 17, 22] in analogy to that expected in asymmetric semiconductor heterostructures.[23-25] This hypothesis is supported by first principle calculations of spin-orbit torques in Pt/Co films[26] but it has also been pointed out that $\mathbf{T}^\perp$ includes the field-like component of the spin torque due to the spin Hall effect (SHE) in the HM layer.[27] On the other hand, $\mathbf{T}^\parallel$ has been shown to be compatible with either the torque due to the spin Hall effect (SHE) in the HM layer[1, 5] or the combination of Rashba effect and spin-dependent conductivity[1]. Although the first experiment on Pt/Co/AlOx suggested that $\mathbf{T}^\parallel$ was too large to be explained by the bulk SHE of Pt alone,[1] later experiments on Ta/CoFeB/MgO showed that the direction of $\mathbf{T}^\parallel$ depends on the sign of the SHE in the HM layer, which is opposite in Ta and Pt.[8] It is now believed that the SHE provides the strongest contribution to $T^{\|}$,[5, 14, 26] but it is still unclear how and to what extent interfacial effects contribute to the generation and absorption of spin currents in HM/FM structures.[6] It is also worth pointing out that contributions from nonlocal (SHE) and interfacial (band structure) effects are not mutually exclusive and lead to equivalent expressions for $T^{\perp}$ and $T^{\|}$, as shown by models based on semiclassical transport,[6, 27-31] quantum kinetic theory[32] and first principle electronic structure calculations.[26]

In addition to the fundamental issues described above, there is still no established consensus on the spin-orbit torque amplitudes as a function of material properties. For example, measurements of $T^{\perp}$ and $T^{\|}$ in Pt/Co/AlOx,[1, 15-17, 33] and Ta/CoFeB/MgO[6, 11] performed by different groups on samples of similar nominal composition and thickness differ significantly from each other. This disparity, due to differences in the preparation and measurement methods, prevents a deeper understanding of spin-orbit torques and makes it difficult to optimize their values for specific material combinations.

In this paper, we report a comprehensive analysis of the current induced spin-orbit torques in Ta(3nm)/CoFeB(0.9nm)/MgO(2.5nm) trilayers, which are representative of the lower electrode of a magnetic tunnel junction device. We investigate the correlation between the spin-orbit torque amplitudes and the structural, electrical, and magnetic properties of as deposited and annealed trilayers in order to elucidate the dependence of $T^{\perp}$ and $T^{\|}$ on material parameters. Adiabatic harmonic Hall voltage measurements show that both torques are very sensitive to changes in the chemical profile of the layers induced by annealing. $T^{\perp}$ and $T^{\|}$ include constant ($T_0^{\perp}, T_0^{\|}$) and magnetization-dependent components ($T_{2,4}^{\perp}, T_{2,4}^{\|}$) that maximize the torque amplitude when the magnetization lies in-plane. The field-like constant component increases in absolute value with increasing annealing temperature, opposite to the antidamping component. In the high current dynamic regime, the effective switching field determined by measuring the probability of magnetization reversal for different external fields increases linearly with the magnetic anisotropy of the layers, differently from $\boldsymbol{T}^\perp$ and $\boldsymbol{T}^\parallel$. This shows that the probability of magnetization reversal depends on several factors, likely related to changes of the sample magnetization and anisotropy due to Joule heating as well as by current- and thermally assisted domain wall nucleation and propagation.

The remainder of this paper is organized as follows. In Sec. II we describe the sample preparation and experimental methods used in this work, namely transmission electron microscopy (TEM) and x-ray absorption spectroscopy (XAS) to characterize the structure and chemical composition of the samples, the anomalous Hall effect (AHE) and x-ray magnetic circular dichroism (XMCD) to measure the magnetization and magnetic anisotropy of CoFeB. We describe also the AC modulation technique used to perform vector measurements of the spin-orbit torques in the limit of small oscillations of the magnetization (low current) and pulsed current injection to induce magnetization switching (high current). The results relative to the sample characterization and equilibrium magnetic properties are presented in Sec. III A and B, respectively. In Sec. IV, we report the measurements of the transverse and longitudinal spin-orbit torque amplitudes as a function of annealing temperature and discuss their relationship with the magnetic properties reported in III. Finally, in Sec. V we study the efficiency of magnetization switching in layers with different magnetic anisotropy fields as a function of current, pulse width, and external magnetic field. Summary and conclusions are presented in Sec. VI.

## II. METHODS

### A. Sample Preparation

Our samples consist of a 0.9 nm thick $Co_{60}Fe_{20}B_{20}$ layer sandwiched between Ta (3 nm) and MgO (2.5 nm), deposited on a thermally oxidized Si wafer by dc magnetron sputtering. Ta was deposited at a rate of 0.15nm/s, $Co_{60}Fe_{20}B_{20}$ was sputtered from a stoichiometric target and deposited at a rate of 0.05 nm/s, and Mg at a rate of 0.1nm/s using an Ar pressure of $2 \times 10^{-3}$ mbar. After deposition, Mg was naturally oxidized in a partial oxygen pressure of 150 mbar for 10 minutes. A 1.6 nm thick Al capping layer was finally deposited and further oxidized (by using plasma oxidation during 85 seconds at $3 \times 10^{-3}$ mbar) on top of the Ta/CoFeB/MgO stack to prevent degradation of the MgO layer in air and during annealing. For magneto-transport measurements, the multilayer film was patterned in the form of a Hall cross [Fig. 2 (a)] with lateral dimensions of 3 µm for the current injection line and 0.5 µm and 1 µm for the Hall measurement line by electron beam lithography and ion beam etching. Unpatterned films diced from the same wafer were used for the TEM and XAS measurements. Annealing of the samples (both patterned and continuous films) was performed in an ultra-high vacuum chamber with a base pressure of $<5 \times 10^{-11}$ mbar at temperatures $T_{ann}$ between 240 and 304ºC, with a dwell time of 30 minutes at each temperature.

### B. TEM and XAS measurements

Transmission electron microscopy (TEM) was performed on as-deposited samples as well as on samples annealed to 240ºC and 296ºC to examine the physical structure of the layers. The images were acquired using a FEI Tecnai F20 high-resolution TEM operated at 200 kV. The samples were prepared in the cross-section geometry by mechanical grinding, precision polishing, and ion milling starting from the continuous film that was used for device fabrication. The annealing of the layers was performed before sample preparation following the same procedure used for the Hall devices.

XAS measurements were performed at the BOREAS beamline of the ALBA synchrotron light source in Barcelona, using 99% circularly polarized light. The x-ray energy was scanned over the $L_{3,2}$ edges of Fe and Co (700-830 eV), and the $K$ edge of Mg (1300 – 1360 eV). The absorption intensity was measured using the total electron yield from the sample and normalized by the electron yield of an Au grid placed after the last optical element of the beamline. All the XAS measurements reported here were carried out at room temperature and at normal incidence. In order to measure the XMCD, a magnetic field of 1 T was applied collinearly to the x-ray beam and consecutive absorption spectra were recorded for parallel or antiparallel alignment of the x-ray helicity and sample magnetization.

**C. Hall effect measurements**

To probe the magnetization in static and dynamic conditions, we have carried out Hall voltage measurements on the devices schematized in Fig. 2 (a). The samples were mounted on a rotatable sample holder, allowing three-dimensional control of the external field direction, and positioned in the gap of an electromagnet producing a magnetic field $B_{ext}$ of up to +/-0.8 T. The Hall measurements presented in this paper are divided into three types according to their purpose.

*1. DC Hall voltage measurements*

Conventional Hall measurements using a low amplitude DC current ($I = 0.1$ mA) were used to probe the response of the magnetization in the presence of an external field. When $I$ is injected along the $x$ direction [Fig. 2 (a)] the Hall voltage is given by:

$$V_H = IR_H = R_{AHE}I\cos\theta + R_{PHE}I\sin^2\theta\sin 2\varphi, \quad (5)$$

where $R_H$ is the Hall resistance, $R_{AHE}$ and $R_{PHE}$ are the anomalous and planar Hall resistances, respectively, and $\theta$ and $\varphi$ represent the polar and azimuthal angle of the magnetization in spherical coordinates. In the above expression we did not include the ordinary Hall effect, which is negligibly small compared to $R_{AHE}$ (4.4-6.1 Ω) and $R_{PHE}$ (0.13-0.18 Ω). When the in-plane field is applied at $\varphi = 0°$ or $90°$ the last term in Eq.1 vanishes and the measured Hall resistance is proportional to the z component of the magnetization, $R_{AHE} \sim m_z = \cos\theta$. By using this simple relation it is straightforward to evaluate the polar angle of the magnetization at the equilibrium position ($\theta_0$, $\varphi_0$) determined by $\boldsymbol{B}_{ext}$ and the uniaxial magnetic anisotropy field ($\boldsymbol{B}_K$) as

$$\theta_0 = \text{acos}\left|\frac{R_H(B_{ext})}{R_{AHE}}\right| \qquad (6)$$

This expression is then used to evaluate $B_K$ by fitting the measurements of $R_H$ versus $B_{ext}$ using a macrospin approximation.

## 2. AC Hall voltage measurements

This second type of measurements consists in the characterization of the current induced spin-orbit torques by detecting the motion of the magnetization vector induced by an AC current.[6, 10, 17] Here we follow the method of Ref.[6], which is appropriate to perform vector measurements of the spin-orbit torques for arbitrarily large angles of the magnetization. Due to the current-induced torques, the injection of a moderate AC current ($I_{AC} = Ie^{i2\pi ft}$) induces small oscillations of the magnetization about its equilibrium position, as schematized in Fig. 2 (b). These oscillations modulate $R_{AHE}$ and $R_{PHE}$ giving rise to a time-dependent Hall voltage signal. The dependence of the Hall voltage on the AC current can be expanded to first order as:

$$V_H(t) \approx V_H(\theta_0, \varphi_0) + \frac{dV_H}{dt}\bigg|_{\theta_0, \varphi_0} = V_H(\theta_0, \varphi_0) + i2\pi fI \frac{dV_H}{dI}\bigg|_{\theta_0, \varphi_0} \qquad (7)$$

Straightforward differentiation of Eq. 5, keeping into account that both $\theta$ and $\varphi$ depend on $I$, gives $\frac{dV_H}{dI} = R_H^f + R_H^{2f}(I)$, where the first and second harmonic Hall resistance components are given by

$$R_H^f = R_{AHE}\cos\theta_0 + R_{PHE}\sin^2\theta_0 \sin 2\varphi_0 \qquad (8)$$

and

$$R_H^{2f} = I(R_{AHE} - 2R_{PHE}\cos\theta_0 \sin 2\varphi_0)\frac{d\cos\theta}{dI}\bigg|_{\theta_0} + IR_{PHE}\sin^2\theta_0 \frac{d\sin 2\varphi}{dI}\bigg|_{\varphi_0}. \quad (9)$$

Here, $R_H^f$ is equivalent to the Hall resistance of conventional DC measurements, whereas $R_H^{2f}$ represents the modulation of the Hall resistance due to current induced effects on magnetization. This dependence can be expressed in terms of the total current-induced effective field $\boldsymbol{B}^I = \boldsymbol{B}^\| + \boldsymbol{B}^\perp + \boldsymbol{B}^{Oe}$, where $\boldsymbol{B}^{Oe}$ is the Oersted field due to the current flowing in the Ta layer. One has

$$\frac{d\cos\theta}{dI} = \frac{d\cos\theta}{d\boldsymbol{B}^I} \cdot \frac{d\boldsymbol{B}^I}{dI} = \frac{d\cos\theta}{dB_\theta^I} b_\theta \quad (10)$$

and

$$\frac{d\sin 2\varphi}{dI} = \frac{d\sin 2\varphi}{d\boldsymbol{B}^I} \cdot \frac{d\boldsymbol{B}^I}{dI} = \frac{d\sin 2\varphi}{dB_\varphi^I} b_\varphi, \quad (11)$$

where $B_\theta^I$ and $B_\varphi^I$ indicate the polar and azimuthal components of $\boldsymbol{B}^I$ and $b_\theta$ and $b_\varphi$ their derivative with respect to the current. The radial component of $\boldsymbol{B}^I$ cannot affect the motion of the magnetization and is thus irrelevant to the discussion of the torques. To measure quantitatively the effective fields $b_\theta$ and $b_\varphi$ by means of Eq. 9, one needs first to calculate the derivatives of $\cos\theta$ and $\sin 2\varphi$ that appear in Eqs. 10 and 11. As the magnetic field dependence of $\cos\theta$ and $\sin 2\varphi$ (proportional to $m_z$ and $m_x m_y$, respectively) is independent of the nature of the field, we can replace $\boldsymbol{B}^I$ by $\boldsymbol{B}_{ext}$ in the derivatives and obtain

$$\frac{d\cos\theta}{dI} = \frac{d\cos\theta}{dB_{ext}} \frac{1}{\sin(\theta_B - \theta_0)} b_\theta, \quad (12)$$

$$\frac{d\sin 2\varphi}{dI} = \frac{d\sin 2\varphi}{dB_\varphi^I} b_\varphi = 2\cos 2\varphi \frac{d\varphi}{dB_\varphi^I} b_\varphi \approx \frac{2\cos 2\varphi}{B_{ext}\sin\theta_B} b_\varphi, \quad (13)$$

where $\theta_B$ is the polar angle defining the direction of $\boldsymbol{B}_{ext}$. The geometrical factor $1/\sin(\theta_B - \theta_0)$ in Eq. 12 accounts for the different projections of $\boldsymbol{B}_{ext}$ and $\boldsymbol{B}^{\perp,\|}$ on the $\boldsymbol{e}_\theta$ axis.

Equation 13 is exact in the case of uniaxial magnetic anisotropy. Using these expressions, $R_H^{2f}$ can be written as

$$R_H^{2f} = I(R_{AHE} - 2R_{PHE}\cos\theta_0 \sin 2\varphi_0) \frac{d\cos\theta}{dB_{ext}}\bigg|_{\theta_0} \frac{1}{\sin(\theta_B - \theta_0)} b_\theta$$
$$+ IR_{PHE}\sin^2\theta_0 \frac{2\cos 2\varphi_0}{B_{ext}\sin\theta_B} b_\varphi. \qquad (14)$$

This expression is valid for arbitrary fields $\mathbf{B}^I$, $\mathbf{B}_{ext}$, and any orientation of $\mathbf{m}$, which makes it possible to measure the vector fields $\mathbf{B}^\perp$ and $\mathbf{B}^\parallel$ as a function of current and position of the magnetization.

Because $R_{PHE}$ is much smaller than $R_{AHE}$ in our samples, the terms proportional to $R_{PHE}$ in Eq. 14 can be neglected, which allows us to determine the polar component of the current-induced field as

$$B_\theta = Ib_\theta = \frac{R_H^{2f}\sin(\theta_B - \theta_0)}{R_{AHE}\dfrac{d\cos\theta}{dB_{ext}}\bigg|_{\theta_0}}. \qquad (15)$$

The general case of non-negligible $R_{PHE}$ has been treated in Ref.[6]. Qualitatively, Eq. 15 states that the current-induced field is proportional to the amplitude of the oscillations of the magnetization normalized by the polar susceptibility of the sample, which is readily evaluated by noting that

$$R_{AHE}\frac{d\cos\theta}{dB_{ext}}\bigg|_{\theta_0} = \frac{dR_H^f}{dB_{ext}}\bigg|_{\theta_0}. \qquad (16)$$

To connect Eq. 15 with the torques in Eqs. 1 and 2 we write the fields $\mathbf{B}^\perp$ and $\mathbf{B}^\parallel$ in spherical coordinates,

$$\mathbf{B}^\perp = -\cos\theta\sin\varphi\,(T_0^\perp + T_2^\perp\sin^2\theta + T_4^\perp\sin^4\theta)\mathbf{e}_\theta - \cos\varphi\,T_0^\perp\mathbf{e}_\varphi, \qquad (17)$$

$$\mathbf{B}^\parallel = \cos\varphi\,(T_0^\parallel + T_2^\parallel\sin^2\theta + T_4^\parallel\sin^4\theta)\mathbf{e}_\theta - \cos\theta\sin\varphi\,T_0^\parallel\mathbf{e}_\varphi. \qquad (18)$$

Measurements of $B_\theta$ performed by setting $\mathbf{B}_{\text{ext}}$ perpendicular to the current ($\varphi = 90º$) therefore give

$$B_\theta = B^\perp = -\cos\theta \sin\varphi \left(T_0^\perp + T_2^\perp \sin^2\theta + T_4^\perp \sin^4\theta\right), \quad (19)$$

whereas measurements performed by setting $\mathbf{B}_{\text{ext}}$ parallel to the current ($\varphi = 0º$) give

$$B_\theta = B^\parallel = \left(T_0^\parallel + T_2^\parallel \sin^2\theta + T_4^\parallel \sin^4\theta\right). \quad (20)$$

Finally, the torque coefficients $T_n^\perp, T_n^\parallel$ are derived by fitting the above expressions for $B^\perp$ and $B^\parallel$ as a function of $\theta$.

The AC Hall voltage measurements presented in this work were performed at room temperature by using a current of amplitude 0.75 mA modulated at $f = 10$ Hz. $V_H$ was recorded during sweeps of the external magnetic field for 10 s at each field step, and fast Fourier transformed to extract $R_H^f$ and $R_H^{2f}$. In Sec. IV, the torque coefficients derived from $R_H^f$ and $R_H^{2f}$ are expressed per unit of magnetic moment, using the same units for torques and effective fields.

In addition to the AHE and PHE, we note that thermal gradients due to Joule heating may introduce additional voltage terms in Eq. 5 due to the anomalous Nernst and spin Seebeck effects. In Pt/Co/AlOx dots, a small Nernst signal due to an in-plane thermal gradient was found and subtracted from the $R_H^{2f}$ data.[6] We find this effect to be entirely negligible in Ta/CoFeB/MgO layers for the ac current density employed here ($<10^7$ A/cm$^2$). Moreover, we have considered the possible effects of a perpendicular thermal gradient. By performing angular-dependent measurements of the anisotropic magnetoresistance, we conclude that such effects, if relevant for the current range employed in this study, have an influence smaller than 5% on the measured $R_H^{2f}$ data. Therefore, we conclude that thermal effects cannot account for the strong variations of the spin-orbit torques as a function of annealing temperature and/or magnetization angle $\theta$ reported in Sec. IV.

### 3. Magnetization switching induced by current pulses

When increasing the current density to about $5 \times 10^7$ A/cm$^2$, the spin torques achieve sufficient amplitude to induce reversal of the magnetization. We have shown in previous work that the switching process can be controlled by adding a small external field parallel to the current direction.[1] The diagram in Fig. 2 (c) exemplifies how this occurs for a perpendicularly magnetized layer. For simplicity, we consider only a torque of the form $\mathbf{T}^\| \sim \mathbf{m} \times (\mathbf{y} \times \mathbf{m})$, which corresponds to an effective field $\mathbf{B}^\| \sim (\mathbf{y} \times \mathbf{m})$ that rotates in the *xz* plane perpendicular to the magnetization. During a current pulse, $\mathbf{B}^\|$ induces a rotation of the magnetization, which can be either assisted or counteracted by applying an external field parallel to *x*, thereby destabilizing or stabilizing the magnetization in one direction (up or down), depending on the sign of the current [Fig. 2 (c)]. Another way to see the same effect is to consider that, when *m* is tilted parallel to *x*, $\mathbf{B}^\|$ has a vertical component that points either up or down depending on the sign of the current.[1] In reality, however, we remark that the switching dynamics is more complex than described above, even in the macrospin approximation. This is because $\boldsymbol{B}^\|$ is not simply proportional to $(\mathbf{y} \times \mathbf{m})$ (Eq. 4) and because the combined action of $\boldsymbol{B}^\|$ and $\boldsymbol{B}^\perp$ should be considered in a time-dependent model.[33, 34]

In Sec. V, we focus on the efficiency of the switching process, independently on the dynamics that leads to magnetization reversal. We have studied the probability of magnetization switching upon the injection of short current pulses (50 ns → 1 μs) as a function of pulse amplitude, duration, and external field applied parallel to the current direction. To ensure the transmission of fast current pulses without significant reflection, we connected a 100 Ω resistance in parallel to the sample and 100 kΩ in series with the Hall voltage arms. $V_H$ was measured before and after each pulse using a DC current of 0.1 mA in order to probe the magnetization.

## III. STRUCTURE AND EQUILIBRIUM MAGNETIC PROPERTIES

### A. Structure

Cross-sectional TEM images of (a-b) as deposited, (c) annealed to 240ºC and (d) 296ºC are shown in Fig. 3. The images reveal that the layers are rather uniform in thickness and homogeneous in the direction parallel to the sample plane. The Ta and CoFeB layers cannot be distinguished due to the small thickness of CoFeB and their similar contrast. Lattice fringes are absent in the Ta/CoFeB region, indicating that it is amorphous, whereas the MgO layer is partially crystalline with (100) texture, as expected for this type of structures.[35, 36] Since both MgO and CoFeB have interplanar spacings around 2.1 Å, it is not easy to discern the presence of crystalline CoFeB grains in the vicinity of MgO, if any. No obvious change of MgO crystallinity is observed by TEM after annealing at 240ºC or 296ºC, consistently with TEM studies of CoFeB/MgO/CoFeB structures grown by rf sputtering and annealed up to 375ºC, which show defective crystallization of the MgO layer.[37]

Figures 4 (a) and (b) show the XAS spectra of Ta/CoFeB/MgO layers annealed at different temperatures measured across the Fe and Co $L_{2,3}$ absorption edges. In the as grown samples, we observe pronounced oxide satellite features on the high energy side of the $L_3$ and $L_2$ resonances for both Fe and Co, which indicate the formation of interfacial Fe-O and Co-O bonds. In agreement with previous findings,[38, 39] the intensity of the oxide multiplet peaks gradually decreases with annealing temperature and finally vanishes at $T_{ann}$ = 296 ºC, indicating that Fe and Co are reduced back to the metallic state upon heating. The Mg $K$ edge spectra shown in Fig. 4 (c) are typical of crystalline MgO, where Mg ions occupy the center of oxygen octahedra.[40] We observe only minor changes of the Mg lineshape from the as grown sample to $T_{ann}$ = 240 ºC, after which the spectra remain the same at least up to $T_{ann}$ = 296 ºC. Such changes, which consist of a sharpening of the first peak at 1312 eV and a variation of the relative intensity of the two peaks at 1318 and 1320 eV, reflect a higher uniformity of the Mg octahedral environment[40] and are attributed to a slight increase of order in the MgO layer.

**B. Magnetic properties**

Figure 5 shows the magnetic field dependence of the Hall resistance of the Ta/CoFeB/MgO layers for $B_{ext}$ applied out-of-plane ($\theta_B = 0°$, a) and nearly in-plane ($\theta_B = 89°$, $\varphi = 0°$, b). The curves are proportional to the $z$-component of the magnetization according to $R_H = R_{AHE} \cos\theta_0$. As inferred from the square loops measured for $\theta_0 = 0°$, all the samples present perpendicular magnetic anisotropy apart from the one annealed to 304°C, which is in-plane. When $B_{ext}$ is applied close to the in-plane direction ($\theta_B = 89°$), the decrease of the Hall resistance is due to the gradual tilt of the magnetization towards the plane. These curves are used to calculate the effective magnetic anisotropy field $B_K$ using a macrospin approximation as $B_K = B_{ext} \left( \frac{\sin\theta_B}{\sin\theta_0} - \frac{\cos\theta_B}{\cos\theta_0} \right)$, where all parameters on the right hand side are determined experimentally. The values of $R_{AHE}$ and $B_K$ obtained from the curves in Fig. 5 (a) and (b) are shown as a function of $T_{ann}$ in Fig. 6 (a) and (b), respectively.

Remarkably, both $R_{AHE}$ and $B_K$ follow a similar bell-shaped curve with a maximum plateau from $T_{ann} = 240$ to 272 °C. $R_{AHE}$ increases from 4.5 Ω in the as-deposited samples to 6.3 Ω for $T_{ann} = 264$ °C, and decreases strongly for $T_{ann} > 272$ °C. This behavior is similar to that of the average magnetic moment of CoFeB ($\bar{m}_{CoFeB}$) derived using the XMCD sum rules [Fig. 6 (a)], indicating that the changes of $R_{AHE}$ mostly reflect variations of the saturation magnetization of the CoFeB layer. The maximum magnetic moments of Fe (1.9 $\mu_B$) and Co (1 $\mu_B$) are attained for $T_{ann} = 272$ °C, consistently with the chemical reduction of the two species observed by XAS. The orbital magnetic moments of Fe (Co) increase from 0.06 $\mu_B$ (0.04 $\mu_B$) in the as grown layers to 0.15 $\mu_B$ (0.11 $\mu_B$) at $T_{ann} = 272$ °C. The values of $\bar{m}_{CoFeB}$, shown in Fig. 6 (a), are obtained by a weighted average of the Co, Fe, and B elemental moments, assuming the latter to be zero. The maximum average moment is thus $\bar{m}_{CoFeB} = 0.98$ μ$_B$ (730 emu), which is smaller with respect to that measured by SQUID for thick amorphous Co$_{60}$Fe$_{20}$B$_{20}$ films (1.65 μ$_B$).[41] We attribute this difference to the reduced magnetic moments of Fe and Co in proximity with Ta.[42] Similarly, we suggest that intermixing of Ta with Fe and Co causes the reduction of the CoFeB magnetization and $R_{AHE}$ observed for $T_{ann}$ > 272 °C [Fig. 6 (a)]. The absence of oxide features in the XAS spectra of Fe and Co for samples annealed to 296 °C (Fig. 4) implies that such a reduction of the magnetic moments cannot be attributed to oxidation. This interpretation is in line with x-ray reflectivity studies of Ta/CoFeB/MgO, which show that the MgO interface remains sharp whereas the Ta interface broadens upon annealing to 300 °C.[43]

$B_K$ is relatively weak in the as-deposited layers (63 mT) and increases upon annealing by a factor ten up to 648 mT for 240 ≤ $T_{ann}$ ≤ 264 °C. For higher annealing temperatures, $B_K$ decreases steeply until at $T_{ann}$ = 304 °C the easy axis rotates in-plane. The coercive field follows a similar behavior, as observed in Fig. 5 (a). We attribute the initial increase of magnetic anisotropy to the increase of crystalline order in the CoFe layer due to the expulsion of B, which is known to take place between 240 and 320 °C and correlate with the coercivity.[44] The decrease of $B_K$ for $T_{ann}$ > 264 °C is likely related to the degradation of the magnetic properties of the sample reflected by the reduced $R_{AHE}$ and magnetic moments [Fig. 5 (a)].

Finally, to contrast the effects of annealing with the electrical properties of the sample, we have measured the longitudinal resistivity of our devices using a four point geometry. The resistivity was found to be 178.5 μΩ·cm for all the annealing steps with ±1.2 μΩ·cm fluctuations depending on the annealing temperature [Fig. 6 (c)]. This suggests that the total thickness of the metallic layers remains constant and supports the conclusion drawn from the XMCD data that the changes of $R_{AHE}$ reflect magnetic rather than electrical contributions.

## IV. SPIN-ORBIT TORQUES

### A. Adiabatic harmonic Hall voltage measurements

The effective field $\boldsymbol{B}^\perp$ was measured using Eq. 19 by applying $\boldsymbol{B}_{ext}$ perpendicular to the current ($\varphi = 90°$). $\boldsymbol{B}^\parallel$ was measured using Eq. 20 by applying $\boldsymbol{B}_{ext}$ parallel to the current ($\varphi = 0°$). To induce significant deviations of $\boldsymbol{m}$ from the perpendicular direction, $B_{ext}$ was applied close to the in-plane direction, but not exactly in-plane in order to keep the sample in a monodomain state.

Figure 7 (a) shows the first harmonic component of the Hall resistance $R_H^f$ measured as a function of $B_{ext}$ by injecting an AC current of 0.75 mA in a Hall bar of Ta/CoFeB/MgO annealed to 240 ºC. Analogously to the DC Hall resistance presented in Fig. 5, $R_H^f$ represents the equilibrium position of the magnetization defined by the angles $\theta_0$ and $\varphi$. Figure 7 (b) shows $R_H^{2f}$ recorded simultaneously with $R_H^f$ for $\varphi = 0°$ (blue open squares) and $\varphi = 90°$ (red solid dots). The two curves have opposite symmetry with respect to $B_{ext}$: for $\varphi = 90°$ $R_H^{2f}$ varies symmetrically with respect to zero field whereas for $\varphi = 0°$ the variation is anti-symmetric. This behavior is associated to the fact that $\boldsymbol{T}^\perp$ is a field-like torque and does not depend on the magnetization direction, whereas $\boldsymbol{T}^\parallel$ does.

Figure 8 (a) and (b) show the polar components of the fields $\boldsymbol{B}^\perp$ and $\boldsymbol{B}^\parallel$ obtained from $R_H^{2f}$ and $R_H^f$ by means of Eq. 15. The data are plotted as a function of $\theta$ for three representative annealing temperatures. The values of the spin-orbit torque coefficients, $T_n^\perp$ and $T_n^\parallel$, are obtained by fitting the curves in Figs. 8 (a) and (b) using Eq. 19 and 20, respectively, and are reported in Figs. 8 (c) and (d). The zeroth order terms $T_0^{\perp,\|}$ represent the amplitude of the usual field-like and antidamping-like spin torques, $\mathbf{y} \times \mathbf{m}$ and $\mathbf{m} \times (\mathbf{y} \times \mathbf{m})$, respectively, whereas the second and fourth order terms $T_{2,4}^{\perp,\|}$ describe anisotropic torque contributions proportional to $\sin^2 \theta$ and $\sin^4 \theta$ (see Eqs. 1 and 2 for $\varphi = 0°, 90°$). The coefficients $T_n^\perp$ and $T_n^\|$, reported in units of mT, represent a torque per unit of magnetic moment. The actual torque acting on the magnetic layer is obtained by multiplication with $\overline{m}_{CoFeB}$ measured by XMCD [Fig. 6 (a)], as shown in Fig. 9 (a).

The data in Fig. 8 evidence several salient features of $\mathbf{B}^\perp$ and $\mathbf{B}^\|$ or, equivalently, of $\mathbf{T}^\perp$ and $\mathbf{T}^\|$. First, both fields are very sensitive to the annealing temperature. Second, at each annealing step, $\mathbf{T}^\perp$ and $\mathbf{T}^\|$ depend strongly on the direction of the magnetization, increasing as $\mathbf{m}$ is tilted in-plane. This anisotropic behavior is described by the $T_{2,4}^\perp$ and $T_{2,4}^\|$ coefficients plotted in Fig. 8 (c) and (d), respectively. Third, the maximum amplitude of the field-like torque $(T_0^\perp + T_2^\perp + T_4^\perp)$ is about five times larger compared to the antidamping torque $(T_0^\| + T_2^\| + T_4^\|)$. Finally, the dependence of $\mathbf{T}^\perp$ and $\mathbf{T}^\|$ on the annealing temperature is remarkably different, as shown in Fig. 9

The behavior of $\mathbf{T}^\perp$ and $\mathbf{T}^\|$ can be analyzed with respect to three approximate temperature regimes corresponding to the as-grown samples, the range between $T_{ann}$ = 240 ºC and 270 ºC where the magnetic anisotropy and magnetization reach maximum values (Fig. 6), and the range beyond 270 ºC where the magnetic properties of the samples degrade, likely due to significant interdiffusion of Ta and CoFeB. These three regions are shown as shaded areas in Figs. 9 (a) and (b). We find that the absolute value of the isotropic field-like torque [solid dots in Fig. 9 (a)] increases by 35% across regions I and II and then by 240% from region II to III, whereas the isotropic antidamping torque [open dots in Fig. 9 (a)] decreases gradually by 40% up to $T_{ann}$ = 288 ºC. The maximum absolute values of $\mathbf{T}^\perp$ and $\mathbf{T}^\|$, obtained for $\theta = 90°$, however, show a rather similar behavior [solid and open triangles in Fig. 9 (a)], decreasing from region I to II and increasing from region II to III, which is opposite to the trend followed by the magnetic anisotropy and magnetization in Fig. 6. Another remarkable feature, shown in Fig. 9 (b), is that the weight of the anisotropic component of $T^\perp$ decreases sharply with the first annealing step from region I to II, remains approximately constant in region II, and decreases further in region III, whereas the anisotropic component of $T^\parallel$ remains constant across regions I and II and increases sharply in region III.

B. Discussion

To explain the spin-orbit torque behavior described above we argue that annealing affects mostly the magnetic layer and its interfaces rather than the bulk properties of the Ta layer. This proposition agrees with the fact that our samples have nearly constant resistivity as a function of $T_{ann}$ [Fig. 6 (c)]. The reduction and partial crystallization of the CoFeB layer going from region I to II, as well as the likely diffusion of Ta into CoFeB[43] and segregation of B at the interface with Ta[45] (region III) might be responsible for the behavior reported in Figs. 8 and 9. Nevertheless, although it is easy to envision that charge scattering, potential gradients, spin reflection and diffusion through the Ta/CoFeB interface shall be affected in such cases, it is difficult to rationalize the spin-orbit torque changes in terms of either the Spin Hall or Rashba models alone. This is because the isotropic and anisotropic amplitudes of the field-like and antidamping-like spin torques present opposite tendencies with respect to annealing [Fig. 9 (a) and (b)]. Such behavior might indicate a different origin for the two torques or opposite tendencies for the real and imaginary part of the spin mixing conductance. We shall note, however, that the common trend observed for the total amplitude of $T^\perp$ and $T^\parallel$ when the magnetization lies in-plane [Fig. 9 (a)] appears to contradict this point. This could be due to the compensation of the isotropic and anisotropic terms for each torque or indicate a similar dependence of $T^\perp$ and $T^\parallel$ on interface properties when the magnetization is in-plane.

We note as well that ab-initio electronic structure calculations for the field-like Rashba torque in Pt/Co/AlOx trilayers predict that the torque magnitude decreases in alloyed Pt/Co interfaces compared to ideal ones,[31] as found also in previous experiments.[6] The Ta/CoFeB/MgO trilayers studied here behave differently, showing that system details, such as chemical composition and mixing of different atomic species, have a strong influence on $\bm{T}^\perp$ and $\bm{T}^\parallel$.

### C. Comparison with spin-orbit torque values reported for Ta/CoFeB/MgO

In order to compare our data with literature values, we consider only the isotropic part of the torques as the existence of anisotropic components has not been measured by other groups. In our samples the isotropic field-like torque per unit of magnetic moment increases in magnitude from $T_0^\perp$ = -2.1 mT/$10^7$Acm$^{-2}$ for the as-grown layers up to -9.5 mT/$10^7$Acm$^{-2}$ at $T_{ann}$ = 288 °C. Kim et al. have found $T_0^\perp$ =-4.1 mT/$10^7$Acm$^{-2}$ in Ta(1.5nm)/Co$_{20}$Fe$_{60}$B$_{20}$(1nm)/MgO(2nm) layers annealed at 300°C for one hour, Suzuki et al. have found $T_0^\perp$=-23 mT/$10^7$Acm$^{-2}$ in Ta(1nm)/Co$_{40}$Fe$_{40}$B$_{20}$(1nm)/MgO(2nm)/Ta(1nm) layers annealed at 300°C for two hours.[22] A more recent study by Zhang et al. have reported $T_0^\perp$ =-2.4 mT/$10^7$Acm$^{-2}$ in Ta(2.5nm)/CoFeB(1nm)/MgO(1.3nm)/Ta(1nm) layers (with unspecified stoichiometry) annealed at 300°C for one hour under an out of plane field of 0.4 T.[11]

For the isotropic antidamping torque, we find $T_0^\parallel$ = -3.2 mT/$10^7$Acm$^{-2}$ for the as-grown layers and -2.0 mT/$10^7$Acm$^{-2}$ for layers annealed up to 288 °C. These values can be converted into an effective spin Hall angle $\theta_{SH}$ by assuming that only the bulk spin Hall effect of Ta contributes to $T_0^\parallel$ according to the following expression: $\theta_{SH} = \frac{2e}{\hbar} M_s t \left[1 - sech\left(\frac{t_{Ta}}{\lambda_s}\right) T_0^\parallel\right]$, where $M_s$ and $t$ are the saturation magnetization and thickness of CoFeB, respectively, $t_{Ta}$ the thickness and $\lambda_s$ the spin diffusion length of the Ta layer. Taking $\lambda_s$ =1.8 nm,[46] we obtain $\theta_{SH}$ = -0.08 and -0.06 for the as-grown and annealed layers, respectively. These values compare with those reported by Kim et al. in Ref. and Zhang et al. in Ref.[11], $T_0^{\parallel}$ = -1.3 mT/$10^7$Acm$^{-2}$, equivalent to a spin Hall angle of about -0.03, and Liu et al. that found $T_0^{\parallel}$ = -3.5 mT/$10^7$Acm$^{-2}$ for Ta(6nm)/CoFeB(1.6)/MgO(1.6) layers of unspecified CoFeB composition annealed to 280 ºC, equivalent to a spin Hall angle of about -0.12. Considering that the Ta resistivity, which is related to the magnitude of the intrinsic spin Hall effect, is very similar in these studies, we believe that the strong dependence on interface and layer structure observed in this work can easily explain the spread of spin-orbit torque amplitudes reported for the Ta/CoFeB/MgO system.

## V. SPIN-ORBIT TORQUE INDUCED SWITCHING

This section is dedicated to determine the switching behavior of the magnetization in the presence of high-amplitude current pulses of sub-µs duration. Due to the high current values both large-amplitude magnetization dynamic effects as well as transient thermal effects become important.[33] Moreover, magnetization reversal in FM/HM layers with lateral dimensions exceeding a few tens of nm proceeds via the nucleation of magnetic domains, which is assisted by joule heating and **$T^{\perp}$**,[16, 17] and domain expansion governed by the combined effects of **$T^{\parallel}$**, the Dzyaloshinskii–Moriya interaction, and external in-plane field.[47-50] Therefore, the action of the spin-orbit torques on the magnetization for high-amplitude current pulses is expected to be different compared to the monodomain, low-current adiabatic regime probed in the previous Section and must be studied separately.

### A. Switching probability

Figure 10 shows two-dimensional plots of the switching probability as a function of pulse amplitude ($I_p$) and in-plane magnetic field applied parallel to the current injection direction [Fig. 2 (c)]. Each diagram was measured by applying 50 ns-long current pulses during a sweep of the in-plane magnetic field. At each field value, a positive and a negative current pulse were applied and the magnetization was measured after each pulse by probing the anomalous Hall voltage. Switching was defined to be successful when the change of the Hall voltage represented more than 90% of the difference between 'up' and 'down' states. The switching probability reported in Fig. 10 depends strongly on the annealing temperature. The switching current threshold is $I_p = 3$ mA for the as-grown layers and about 4 mA for the annealed samples. Note that switching can be measured only within the field boundaries defined by the coercivity of the layers, since the magnetization is probed at equilibrium. Since the spin-orbit torque that induces the reversal of the magnetization is $\boldsymbol{T}^{\parallel}$,[1] switching is either assisted or counteracted by the in-plane external field when $\boldsymbol{B}^{\parallel} \sim (\boldsymbol{y} \times \boldsymbol{m})\, T_0^{\parallel}$ is either parallel or antiparallel to it. This explains why, at the threshold current, the external field required for switching is maximum and coincides with the coercive field of the sample in the absence of pulse injection. The shape of the switching regions (red hatched areas in Fig. 10) also reveals details about the mechanisms that determine the reversal process. In the as-grown as well as annealed layers, there is a region where the minimum external field required for switching decreases linearly with increasing $I_p$, reflecting a proportional increase of $\boldsymbol{T}^{\parallel}$ and therefore a higher switching rate. In the as-grown layers, on the other hand, the outer boundary of the switching region remains approximately constant and equal to the coercivity until $I_p$ reaches 5 mA, whereas in the annealed layers we observe a reduction of the outer boundary field that becomes more pronounced with increasing annealing temperature. We believe that this tendency originates from the changes of magnetic anisotropy induced by annealing, which result in a stronger but also more temperature-sensitive $B_K$, and therefore to a stronger reduction of the coercivity during the current pulses.

**B. Switching versus external field and magnetic anisotropy**

Figure 11 (a) shows the minimum external field required to achieve magnetization reversal as a function of $I_p$. The curves are obtained by taking horizontal cuts of the diagrams in Fig. 10 for $4 < I_p < 6$ mA. We recall that the (in-plane) external field serves two purposes in the switching process: the first is to break the symmetric effect of $\boldsymbol{T}^{\parallel}$ with respect to states where the magnetization points either up or down;[1, 7] the second is to unwind the chirality of domain walls imposed by the Dzyaloshinskii–Moriya interaction and induce a component of the magnetization parallel to the current to allow for the expansion of a reversed domain under the action of $\boldsymbol{T}^{\parallel}$.[49] The external field is therefore an important parameter to induce deterministic magnetization reversal and define the working conditions of spin-orbit torque devices in applications. For a given current, Fig. 11 (a) shows that the minimum field required to reliably switch the magnetization increases with the magnetic anisotropy of the CoFeB layers and is maximum in samples annealed between 248 and 264 ºC. This is not surprising because the field-induced tilt of the magnetization along the current direction (in the small angle limit) is inversely proportional to $B_K$. However, we also find that the minimum field decreases much faster as a function of current in the hard layers compared to the soft layers. The slope of the curves in Fig. 11 (a) gives an effective field per current, which we call $b^*$, that describes this behavior. Figure 11 (b) shows $b^*$ as a function of $T_{ann}$. We measure $b^* = 17$ mT/$10^7$ A/cm$^2$ in the as-grown layers, increasing up to $b^* = 125$ mT/$10^7$ A/cm$^2$ in samples annealed to 248 ºC. A larger $b^*$ implies a faster decrease of the minimum external field as a function of current. Moreover, we observe that $b^*$ is proportional to $B_K$ [inset of Fig. 11 (b)], meaning that the effect of the current is comparatively larger in hard magnetic layers relative to soft layers. This result is tentatively attributed to a stronger temperature-induced loss of the magnetic anisotropy in the hard layers or to a decrease of the domain wall pinning energy with annealing temperature.[49] Note that the resistivity remains constant throughout all the annealing steps [Fig. 6 (c)], so that the sample temperature reached during pulses of identical amplitude and duration is expected to be approximately the same.

**C. Energy considerations and switching efficiency**

The minimum energy required to switch the magnetization was determined by measuring the critical switching current ($I_c$) for square pulses of width ranging from 75 ns to 1 µs at each annealing stage. The external field was fixed at 10 mT and $I_c$ was defined as the minimum current that switched the magnetization 10 times over 10 consecutive trials without failure. Figure 12 (a) shows $I_c$ as a function of pulse width. We observe an exponential decay of $I_c$ for all samples, which is expected based on spin transfer torque models of magnetization switching in the thermally activated regime.[51] We also notice that $I_c$ depends strongly on the annealing conditions: as-grown layers switch at lower current compared to annealed ones. This is consistent with the switching diagrams reported in Fig. 10 and due to the increase of magnetic anisotropy energy ($E_m = \bar{m}_{CoFeB} B_K$). However, we find that the threshold switching current in samples annealed to 240 – 264 ºC increases only by 30% with respect to the as-grown layers, similar to $\bar{m}_{CoFeB}$, whereas $B_K$ increases by ten times. In order to compare the switching efficiency in different samples, we have calculated the ratio between the electrical pulse energy ($E_p$) required for switching and $E_m$ for each data point, as given by $E_p/E_m = I_c^2 R_s \tau_p / \bar{m}_{CoFeB} B_K$. The smaller this ratio is, the larger is the efficiency of the current to switch the magnetization for a given magnetic anisotropy energy. In Fig 12 (b) we present the ratio $E_p/E_m$ as a function of $\tau_p$ and annealing temperature, where light areas represent higher efficiency (lower $E_p/E_m$ ratio) and darker areas lower efficiency (higher ratio). We observe that the energy efficiency is lowest in the as-grown layers and maximum for $T_{ann}$ = 240-280ºC, similar to $B_K$ [Fig. 6(b)] and $b^*$ [Fig. 11(b)]. A similar behavior is found if, instead of the magnetic anisotropy energy, one considers the magnetic energy stored in the layers, given by the product of $\overline{m}_{CoFeB}$ with the coercive field measured at different annealing temperatures.

## VI. CONCLUSIONS

In summary, we have studied the effect of thermal annealing on the current induced spin-orbit torques in perpendicularly magnetized Ta/CoFeB/MgO layers. We have found that both the field-like and antidamping torques have a complex behavior depending on the annealing history of the sample and the magnetization direction.

Measurements of the field-like and antidamping spin-orbit torques in the adiabatic regime show that both include a constant and a magnetization-dependent component, the amplitude of which is maximum when the magnetization lays in-plane. The absolute value of the constant field-like component ($T_0^\perp$) increases with increasing annealing temperature, opposite to the antidamping component ($T_0^\parallel$). On the other hand, the relative weight of the magnetization-dependent field-like component $[(T_2^\perp + T_4^\perp)/T_0^\perp]$ is maximum in the as-grown layers, whereas the magnetization-dependent antidamping component $[(T_2^\parallel + T_4^\parallel)/T_0^\parallel]$ is constant up to $T_{ann}$ = 264ºC and increases above this temperature. The maximum amplitude of the field-like torque is about a factor five larger compared to the antidamping torque at all temperatures. This behavior is related to the chemical and structural changes taking place in the samples, which also induce significant variations of the magnetic moment and anisotropy of the layers, as evidenced by XAS, XMCD, and Hall effect measurements. The magnetic moment and magnetic anisotropy of Ta/CoFeB/MgO increase, respectively, by 36% and 1000% in samples annealed to 264ºC and decrease afterwards by 30% and 50% when annealed to 296ºC. These changes are attributed to the reduction and partial crystallization of the CoFe layer and, for annealing temperatures above 270ºC, to intermixing of Ta and CoFeB.

Although the variations of the spin-orbit torque magnitude and magnetic properties appear to be correlated, an explanation of this behavior remains to be given.

Measurements of the switching probability in the high current dynamic regime show that more current is required to reverse the magnetization of hard magnetic layers compared to soft ones. However, the ratio between the energy delivered by each current pulse and the magnetic anisotropy energy has a minimum in samples with high magnetic anisotropy, implying that switching becomes relatively more efficient in harder magnetic layers. This behavior may be due to transient joule heating effects, which can affect the spin-orbit torques, and activation energy for magnetization reversal, including domain wall depinning, in different ways.

Ta/CoFeB/MgO layers are widely employed to fabricate perpendicular magnetic tunnel junctions.[4, 52] Post-deposition annealing of such tunnel junctions is critical to induce the crystallization of the initially amorphous CoFeB layer, optimize the perpendicular magnetic anisotropy, and increase the tunneling magnetoresistance. In the case of the thin $Co_{60}Fe_{20}B_{20}$ layers employed in this study, moderate annealing to about 260°C appears to be the best choice to maximize the magnetic anisotropy, magnetization, and spin-orbit torque switching efficiency. This temperature is close to that used to optimize the tunneling magneto-resistance in Ta/CoFeB/MgO/CoFeB/Ta tunnel junctions where the thickness of the free layer is about 1nm.[4]

## VI. ACKNOWLEDGMENTS

This work was supported by the the European Commission under the Seventh Framework Programme (GA 318144, SPOT), the European Research Council (StG 203239 NOMAD), the Ministerio de Economía y Competitividad (MAT2010-15659), and the Swiss

Competence Centre for Materials Science and Technology (CCMX). The XAS experiments were performed at the BOREAS beamline of the ALBA Synchrotron Light Facility.

**FIGURES**

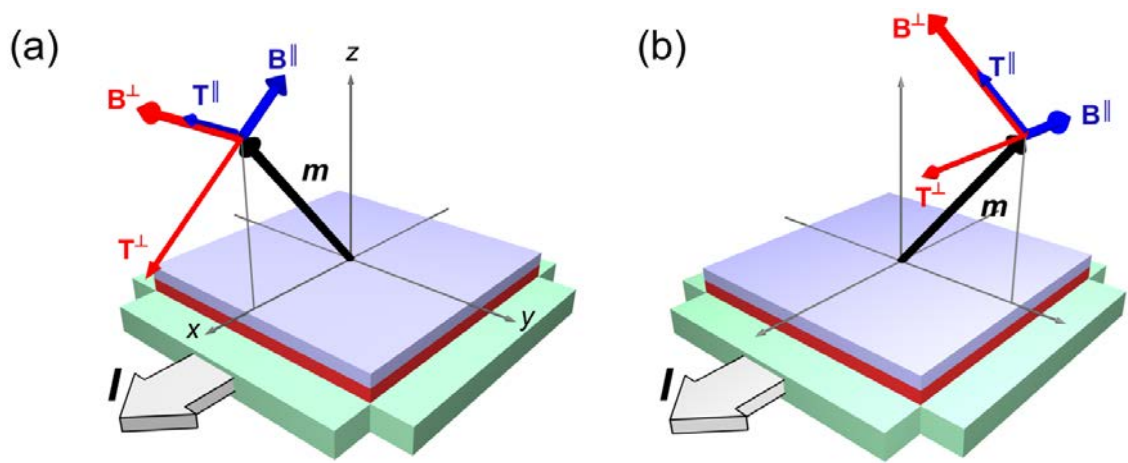

FIGURE 1. Spin-orbit torques $\boldsymbol{T}^{\perp}$ and $\boldsymbol{T}^{\parallel}$ and corresponding effective fields $\boldsymbol{B}^{\perp}$ and $\boldsymbol{B}^{\parallel}$ when the magnetization is tilted parallel to the current direction (a) and perpendicular to it (b). The length of the arrows is proportional to the intensity of the torques and fields measured for as grown Ta/CoFeB/MgO trilayers.

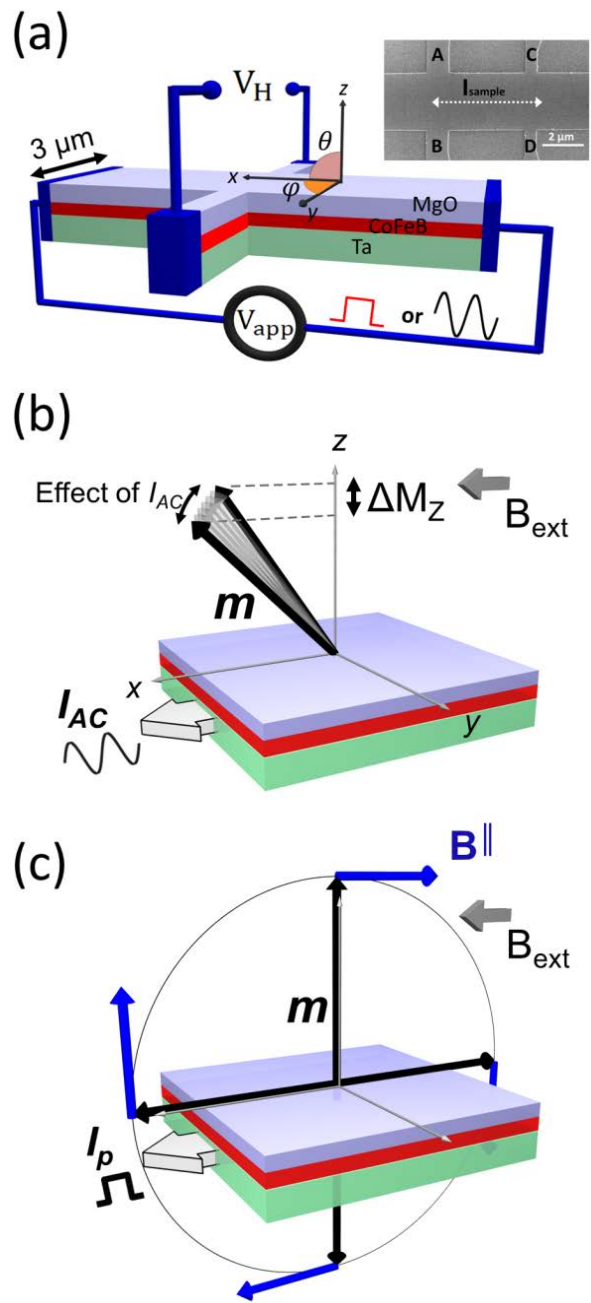

FIGURE 2. (a) Schematic of the sample and measurement geometry. Inset: Scanning electron micrograph image of a sample. (b) Small oscillations of the magnetization induced by an AC current. (c) Switching of the magnetization induced by the combined action of $B^{\parallel}$ and the external field.

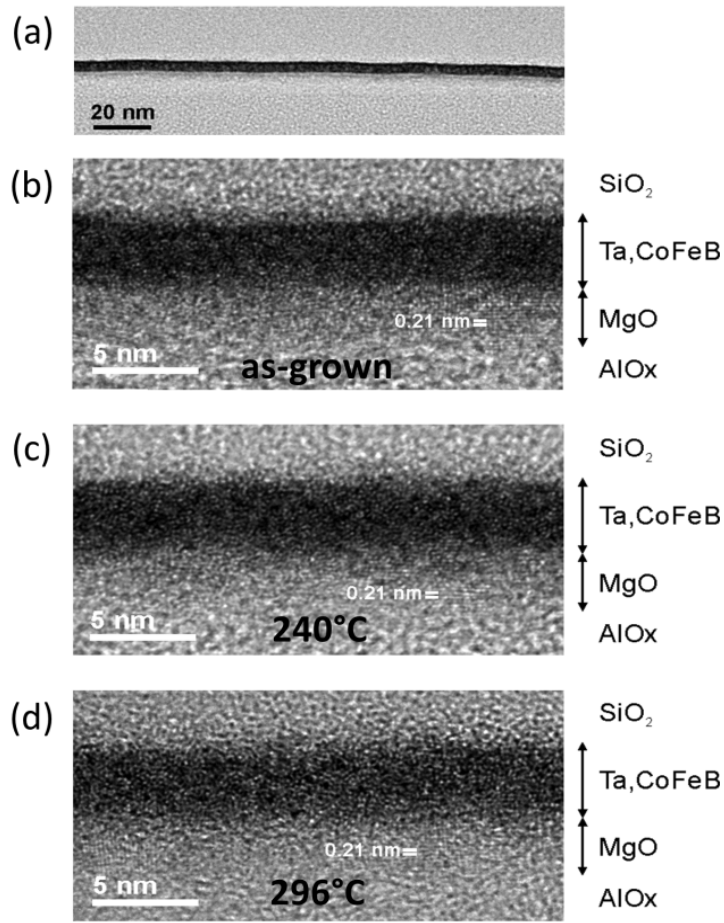

FIGURE 3. Cross-sectional TEM images of the samples annealed to different temperatures: (a) and (b) as grown, (c) $T_{ann}$ = 240 ºC, (d) $T_{ann}$ = 296 ºC.

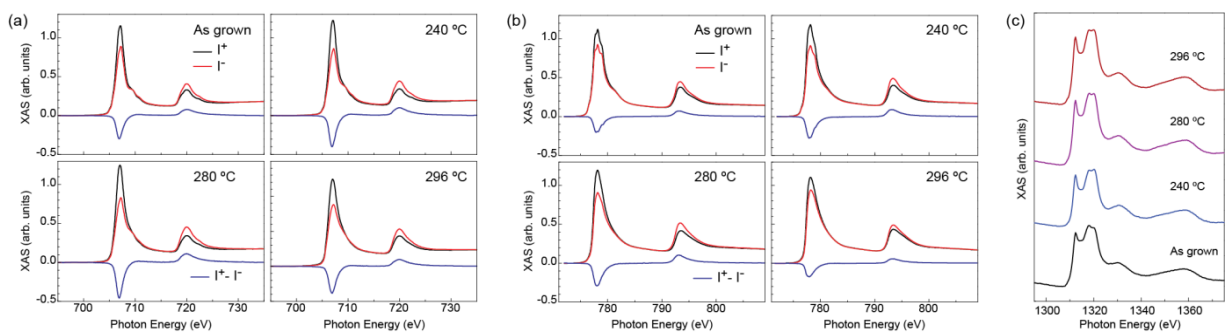

FIGURE 4. X-ray absorption and magnetic circular dichroism spectra of as grown and annealed Ta/CoFeB/MgO trilayers measured at the Fe $L_{2,3}$ edges (a), Co $L_{2,3}$ edges (b), and Mg $K$ edge (c). The spectra were recorded at normal incidence at room temperature in a magnetic field of 1 T.

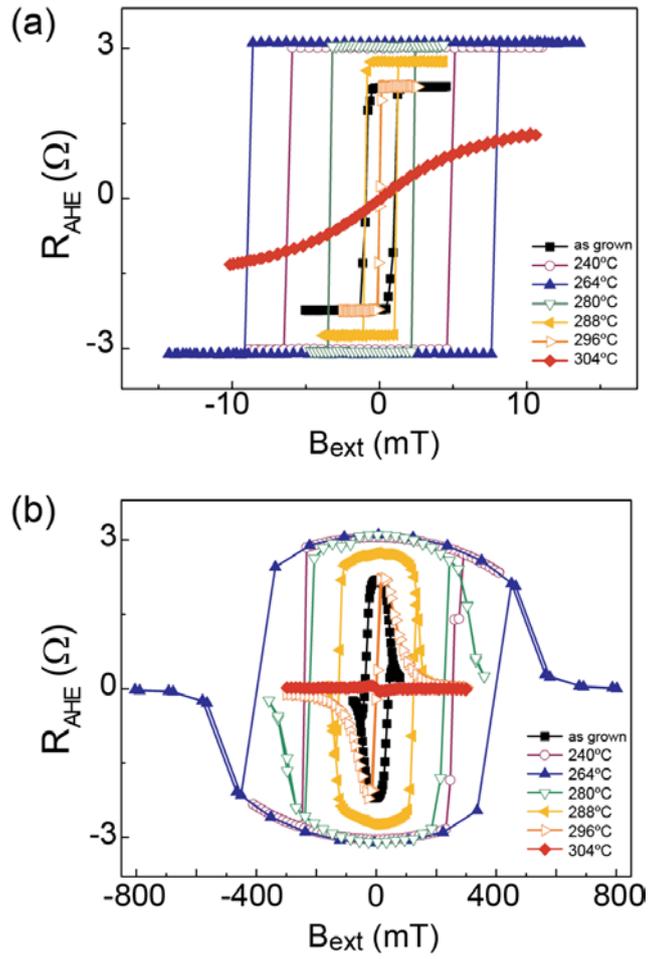

FIGURE 5. Anomalous Hall resistance measured on as grown and annealed Ta/CoFeB/MgO trilayers with the external field applied (a) out-of-plane ($\theta_B = 0°$) and (b) in-plane ($\theta_B = 89°$).

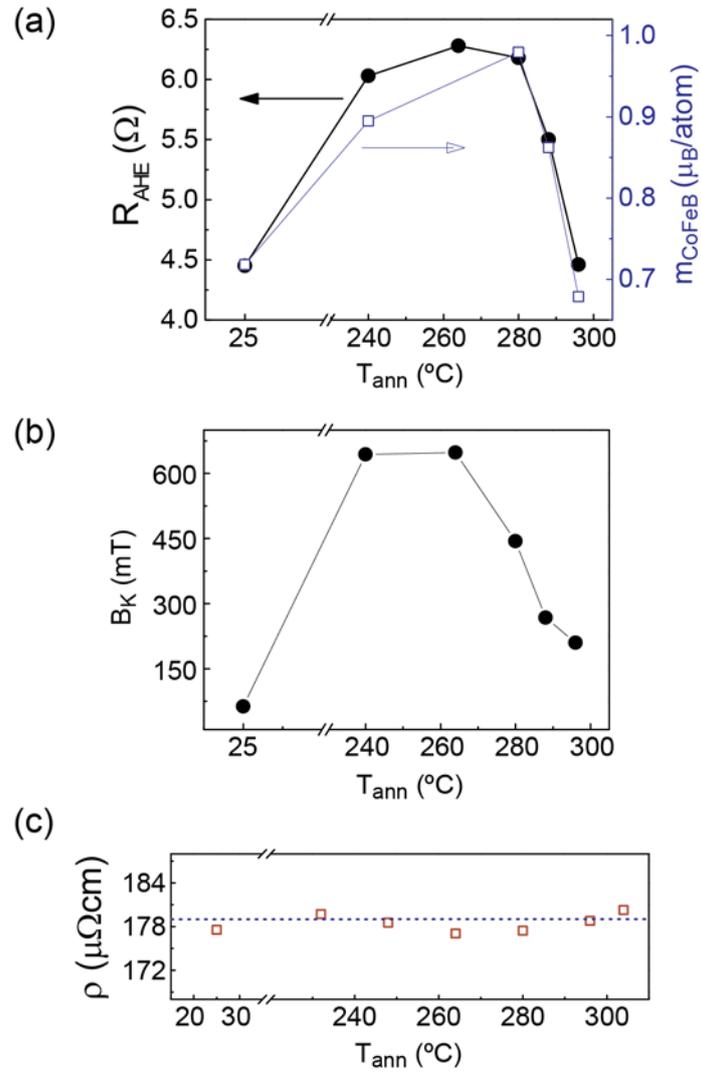

FIGURE 6. (a) Anomalous Hall resistance and average magnetic moment per atom measured by XMCD as a function of annealing temperature. (b) Perpendicular anisotropy field calculated from the curves shown in Fig. 5 (b). (c) Longitudinal resistivity of Ta/CoFeB/MgO as a function of annealing temperature.

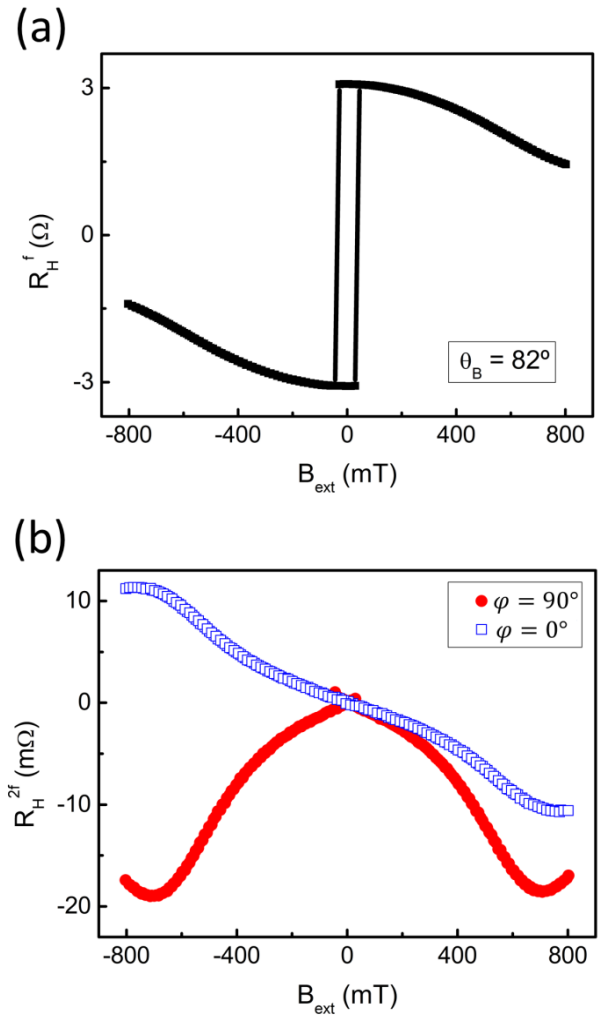

FIGURE 7. (a) First and (b) second harmonic Hall resistance for the sample annealed at 240˚C. The external field has been applied at $\theta_B = 82°$. Measurements of $R_H^f$ at $\varphi = 90°$ and $\varphi = 0°$ give very similar curves due to the in-plane symmetry of the layers.

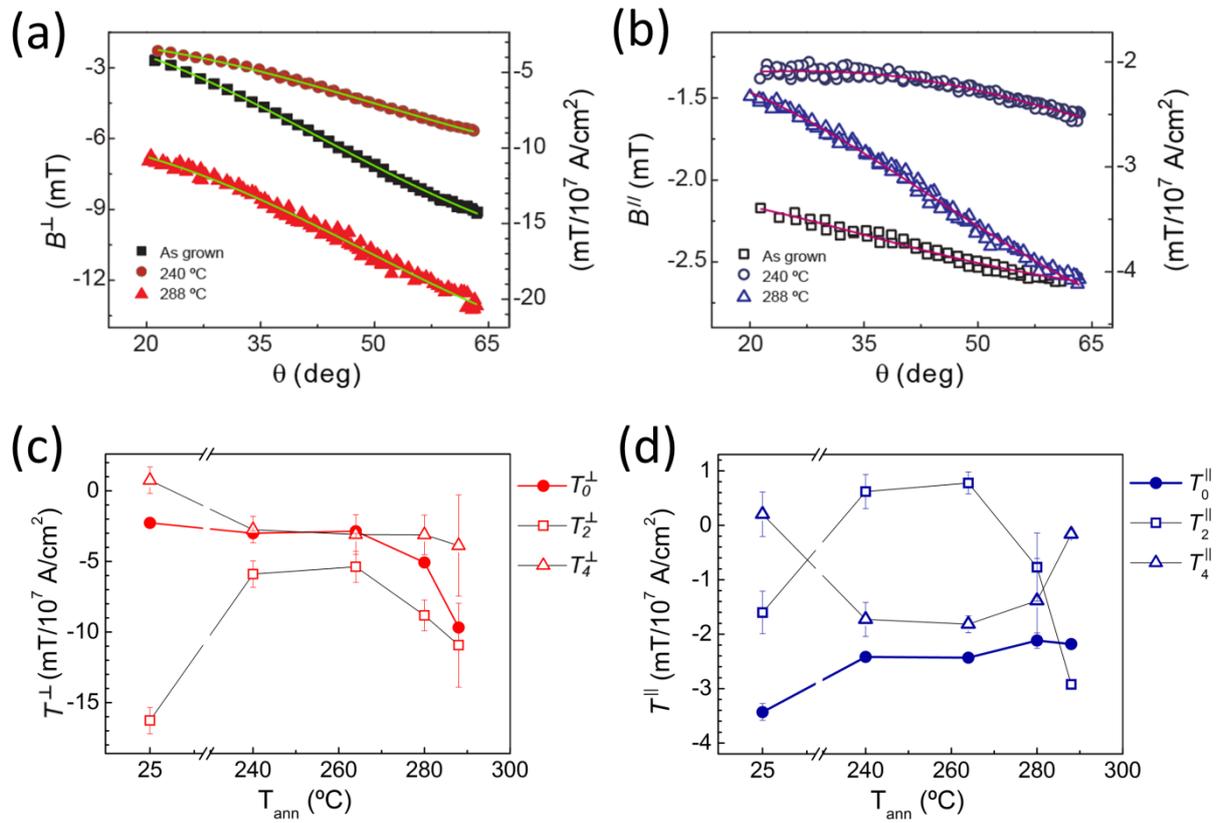

FIGURE 8. Current-induced effective fields $B^\perp$ (a) and $B^\parallel$ (b) as a function of the polar angle of the magnetization. The scales on the left correspond to an injected current of 0.75 mA; the scales on the right gives the field values normalized to a current density of $10^7$ A/cm$^2$. The solid lines are fits to the data according to Eqs. 19 and 20. (c) Spin-orbit torque coefficients $T_n^\perp$ and (d) $T_n^\parallel$ as a function of annealing temperature. The values in (c,d) are normalized to a current density of $10^7$ A/cm$^2$.

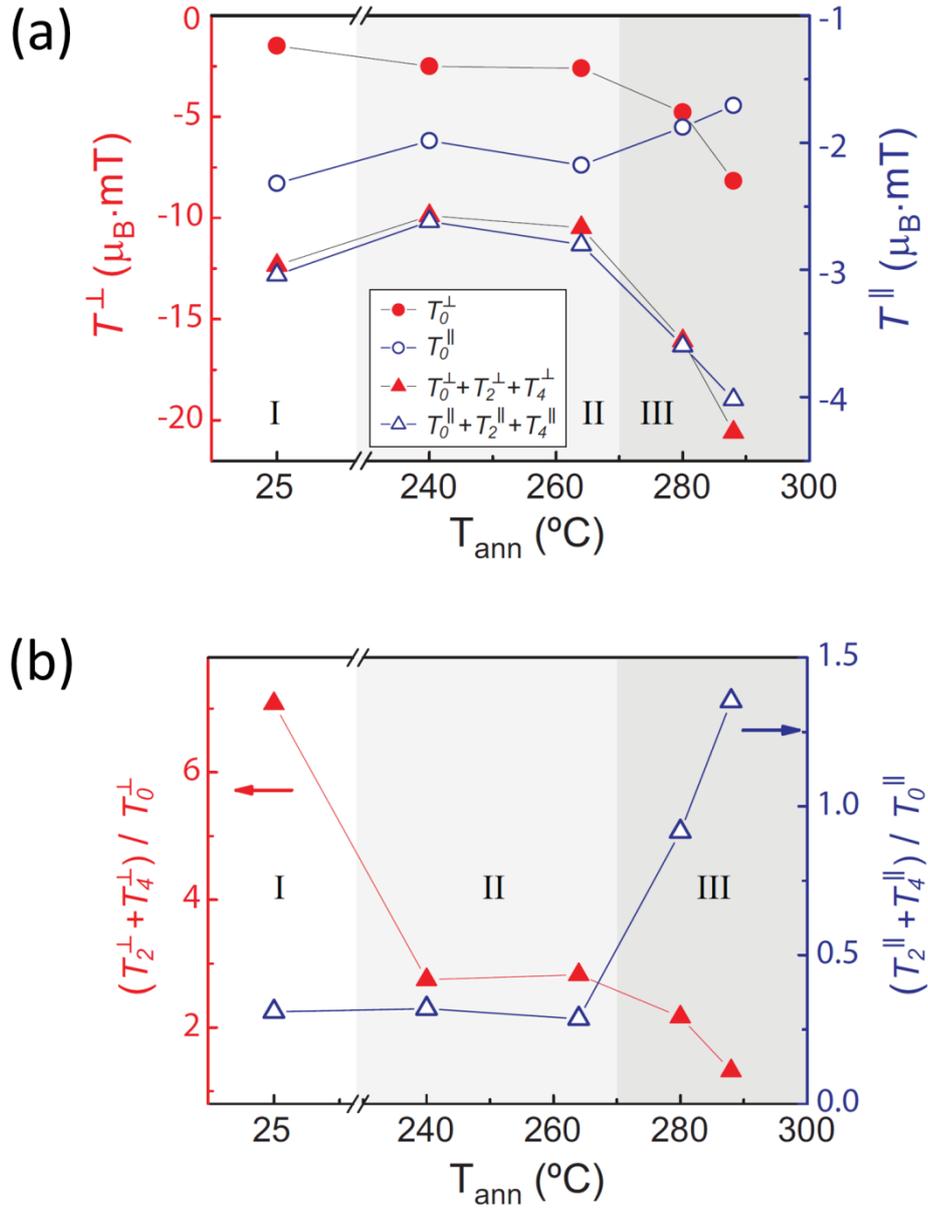

FIGURE 9. (a) Isotropic (circles) and maximum torque amplitudes (triangles) obtained by multiplying the torque coefficients in Fig. 8 (c) and (d) by the average magnetic moment of CoFeB. (b) Ratio between the anisotropic and isotropic torque amplitudes.

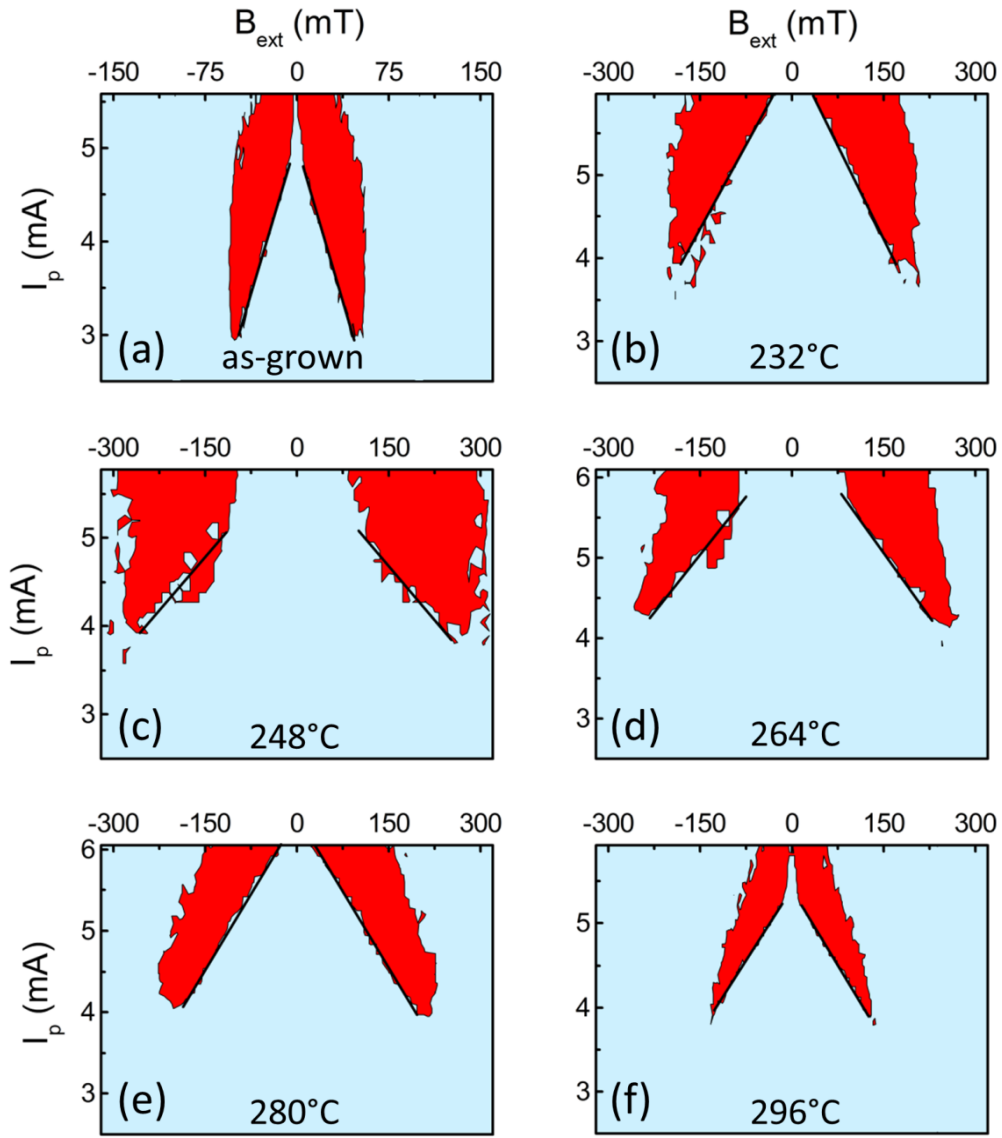

FIGURE 10. Switching diagrams of as-grown (a) and annealed (b-f) Ta/CoFeB/MgO trilayers, showing the difference between the anomalous Hall voltages measured after consecutive positive and negative current pulses as a function of $I_p$ and $B_{ext}$. Red/light blue colors represent switching/non-switching regions. Black lines are guide to the eyes emphasizing the linear behavior of the minimum external field required for switching as a function of $I_p$.

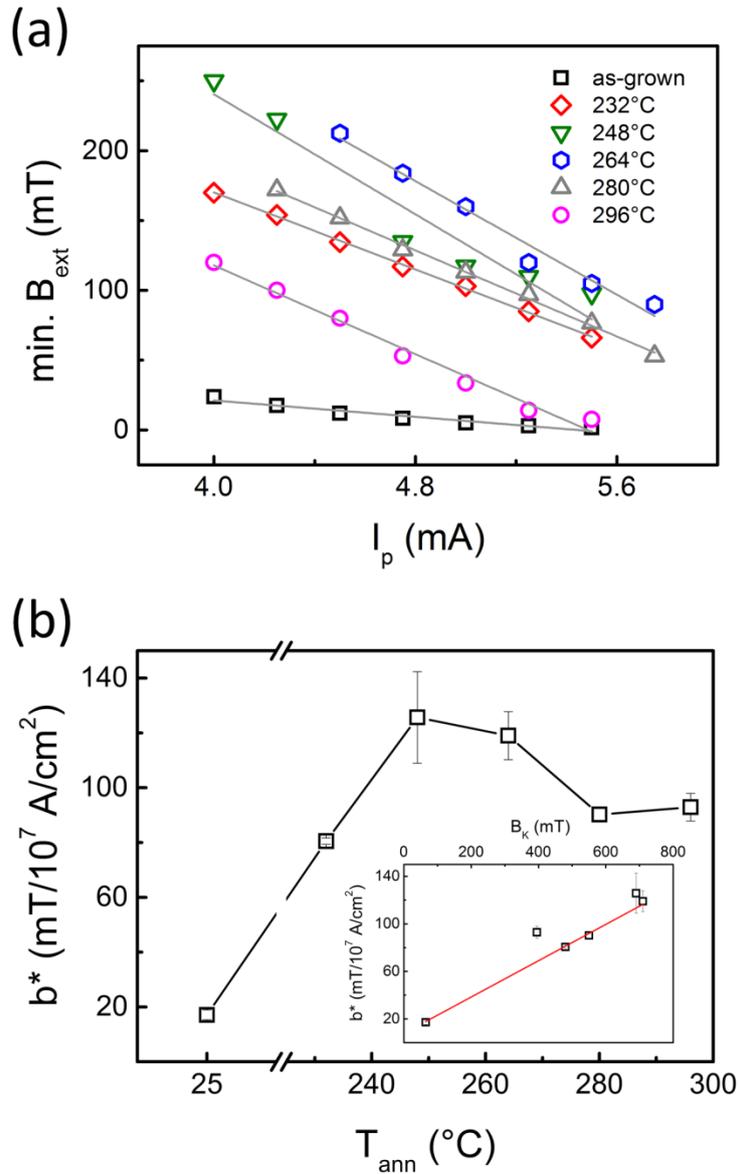

FIGURE 11. (a) Minimum external field required for switching as a function of current pulse amplitude. (b) Effective switching fields normalized to $10^7$ A/cm$^2$, determined from the slopes of the linear fits in (a), as a function of the annealing temperature of the layers. Inset: the same data plotted as a function of the effective magnetic anisotropy field.

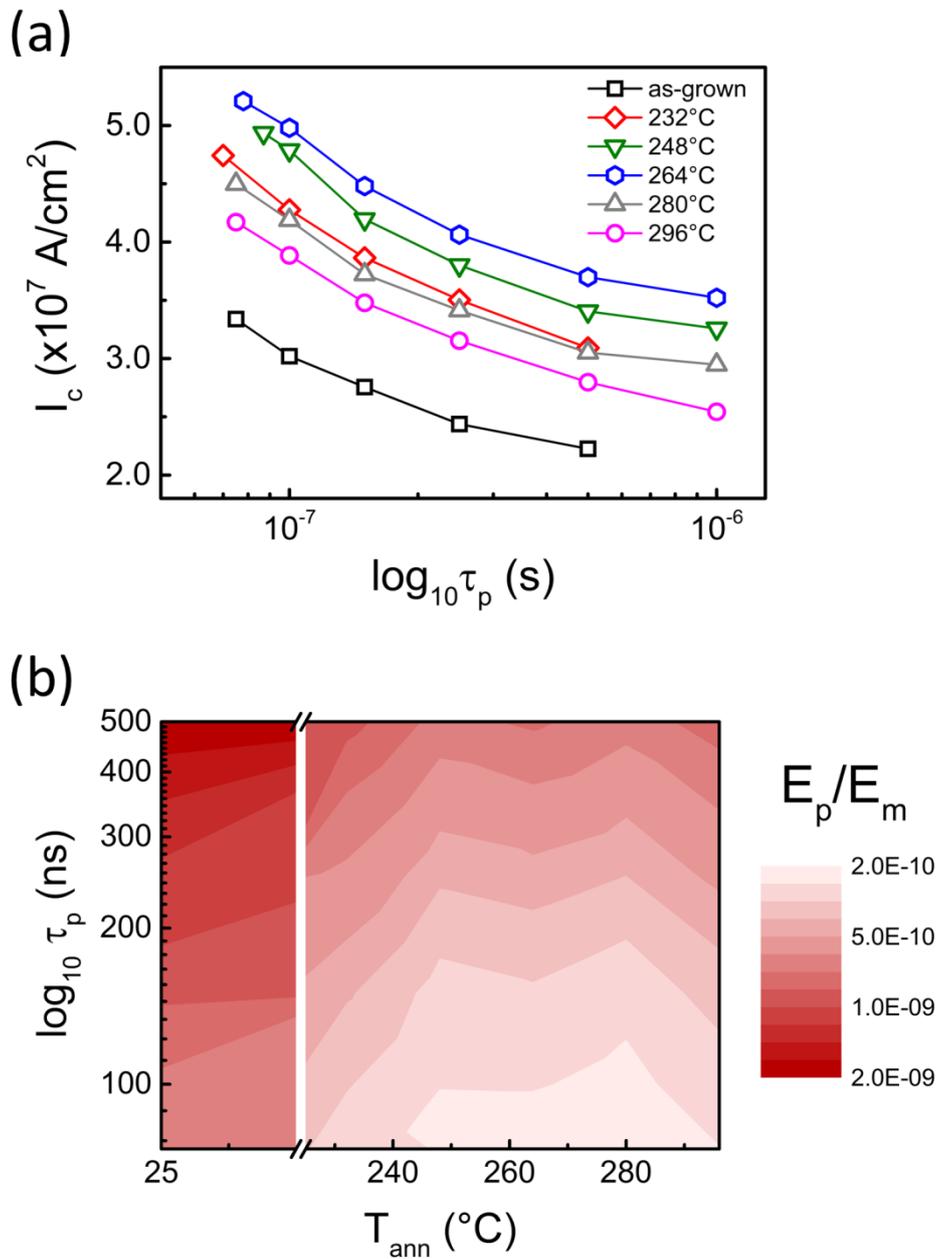

FIGURE 12. (a) Critical switching current density as a function of the pulse width. (b) Switching efficiency diagram as a function of the annealing temperature and current pulse width. The color scale represents the ratio between the electrical pulse energy ($E_p$) and magnetic energy ($E_m$).